# Phototransistor Behavior Based on Dye-Sensitized Solar Cell

X. Q. Wang[a], C. B. Cai, Y. F. Wang, W. Q. Zhou, Y. M. Lu and Z. Y. Liu,

Physics Department, Shanghai University, Shanghai 200444, China

**Abstract:**

In the present work, a light-controlled device cell is established based on the dye-sensitized solar cell using nanocrystalline $TiO_2$ films. Voltage-current curves are characterized by three types of transport behaviors: linear increase, saturated plateau and breakdown-like increase, which are actually of the typical performances for a photo-gated transistor. Moreover, an asymmetric behavior is observed in the voltage-current loops, which is believed to arise from the difference in the effective photo-conducting areas. The photovoltaic voltage between the shared counter electrode and drain ($V_{CE-D}$) is investigated as well, clarifying that the predominant dark process in source and the predominant photovoltaic process in drain are series connected, modifying the electric potential levels and thus resulting in the characteristic phototransistor behaviors.

With the low-cost and lightweight expectation, organic phototransistors have been attracted a great deal of interest in recent years[1-3]. For the operation with light, a back gate electrode is usually introduced to incite the separation of the photo-excited electrons and holes in bulks[4-6]. This is called the charge separation process. In the case of solar cells, this process may emerge either from the inner electric field such as in organic heterojunction solar cells (OHSC), or from the fast electron injection and subsequent redox at the interfaces such as in dye-sensitized solar cells (DSSC)[7-9]. These imply that a photoconductor based on solar cells may perform like a phototransistor without a back gate.

For the OHSC based device, the charge transfer dynamics (CTD) may be fluctuated as the space charges are subject to the applied voltage[10]. For the DSSC-based devices, however, CTD is probably ascertained as the charge transport is dominated by the diffusion, i.e., as a result of carrier concentration gradient. It is thus supposed that the DSSC-based phototransistor may perform more readily. Moreover, a few advantages can be expected for the DSSC-based devices, including the application of wide band-gap semiconductors insensitive to background thermal noise[11], and the flexibility of spectral response by the choice of the proper dye[11, 12].

In the present work, a phototransistor is established based on the DSSC with the nanocrystalline $TiO_2$ films. The transistor characteristics of the tentatively designed device under the illumination of a sun simulator will be studied in details, and will further be discussed in view of the electric potential level diagram.

Before a 5×5 mm² monolayer of $TiO_2$ nanoparticles was prepared by screen printing on the F-doped transparent conducting glass (FTO), two separated electrodes drain and source had been introduced on FTO by a simple electric-burning method, as shown in Fig.1 (a). This method employs one electrode connecting with the edge of FTO, and the other with a needle scratching along the middle of FTO. Under a given voltage, an approximate 100 μm insulating groove was drawn.

Fig.1 (b) shows the configure of a transistor-like structure consisting of the counter electrode (CE) opposite to the source (S) and drain (D), as well as the involved $TiO_2$ nanoparticles near the insulating regions. The device cell was sealed by a 25 μm-thick plastic spacer together with the platinized CE after sensitizing for 20 h in the dye of N719. The resultant device was filled with the electrolyte including 0.1 M LiI, 0.1M $I_2$, 0.5 M 4-tert-butyl pyridine, and 0.6 M 2-Dimethyl-3-propylimidazolium iodide in acetonitrile. The voltage-current curves are measured by Keithley 2420 and 2100 digital meters. An AM1.5 light was provided by a commercial solar simulator (SAN-EI XES-151S) equipped with a 150W Xenon lamp. The photovoltaic characteristics of both CE-Drain (CE-D) and CE-Source (CE-S) are similar. The short-circuit current density ($J_{SC}$) and the open-circuit voltage ($V_{OC}$) are approximately 9 mA/cm$^2$ and 0.64 V, respectively. The efficient is achieved 3.9 %.

Figure 2 shows the *V-I* curves of Drain-Source (DS) under two different conditions: illumination (on state) and dark (off state). In the case of on state, the plot is characterized by three types of behaviors: a linear dependence of the response current ($I_{DS}$) at low voltage ($V_{DS}$), a gradual transition to the saturation at moderate $V_{DS}$, and a breakdown-like surge at high $V_{DS}$. These are actually of the typical performances of a phototransistor. A calculated *dI/dV* is illustrated in the inset of Fig. 2. In the range of 0.3~0.7 V, the stably zero differential conductance suggests two significant characteristics for the potential application of the present device, low power and high stability. In the case of off state, $I_{DS}$ is characterized with a nearly zero value, as $V_{DS}$ is smaller than 0.7 V.

It is generally believed that under illumination, electrons of dye molecules are excited from the ground state (S) to the higher excited state (S*), and then inject into the conduction band (CB) of TiO$_2$ particles, leaving the dye molecules to an oxidized state (S$^+$). If one supposed that the excess photo-excited electrons in TiO$_2$ particles might transport transversely and might form the response current $I_{DS}$ under biases, then the photo-induced $I_{DS}$ in the present device would not be related to the I$^-$/I$_3^-$ redox, but would be subject to the cross-section area of TiO$_2$ on the groove. This assumption, however, doesn't agree with the experimental observations which indicate that the $I_{DS}$ is dependent on both the redox and the effective area on the electrode.

As shown in Fig. 2, an off-state-like behavior appears under illumination, which was intentionally prepared by using the pure acetonitrile as the electrolyte solution instead of the normal solution containing $I^-/I_3^-$ redox. This implies that $I^-/I_3^-$ redox actually plays a crucial role in the final transport performance, rather than photo-excited process only.

To address the issue regarding the area dependences, $V$-$I$ loops are measured by scanning voltage: 0V→1.1V→0V→-1.1V→0V. Note that the loop of current density ($J_{DS}$) versus voltage is derived from a simple calculation of $I_{DS}$ being divided by the effective area of $TiO_2$ on the electrode where a positive voltage is applied. If $I_{DS}$ was determined by the cross-section area, the $I_{DS}$ (left axis)-$V_{DS}$ loop would be symmetric. However, in the region II of Fig. 3, an asymmetric characteristic of the loop can be seen. In contrast, the $J_{DS}$ (right axis)-$V_{DS}$ loop appears symmetric clearly, implying a relationship between the effective area of $TiO_2$ on the electrode and the response current in the saturation region.

The above phenomena are understandable if one considers the present device as two separated DSSCs connected with a common CE. The transport process can be divided into two parts: photovoltaic process (PP) and dark process (DP) in both drain and source. If the potential difference between $TiO_2$/dye levels and redox level reach $V_{OC}$, PP and DP will be counterbalanced. The increase or decrease in the potential difference will enhance or depress DP correspondingly, but do nothing to PP which is dominated by the illumination conditions.

To ascertain how the potential difference varies with the bias, the photo-excited voltage ($V_{CE-D}$) is examined during the loop measurement, as seen in Fig. 4. It is revealed that at $V_{DS}>0$, $V_{CE-D}$ decreases and the slope is approximate to −1, and at $V_{DS}<0$, $V_{CE-D}$ shows a step in region II and a little increasing in both regions of I and III, similar to the performance of $I_{DS}$ showed in Fig. 3. For a clear illustration, $V_{CE-D}$ (left axis) in the rectangle frame is enlarged and replotted in Fig. 4 (b), together with the $I_{DS}$ (right axis) partly extracted from Fig. 3.

Figure 5 illustrates a general diagram for various potential levels probably existing in the present device. The photovoltaic process is counterbalanced by the dark process in both drain and source at $V_{DS}=0$. As $V_{DS}$ increases, the TiO$_2$/dye levels on source ($L_S$) move upwards relative to $L_R$, while $L_D$ goes down. The former will enhance the dark current around the source (black arrow), while the latter will depress the DP around the drain and maintaining the photo-excited current (orange arrow). A competitive consequence is that $I_{DS}$ increases linearly first and then reaches saturation value as $V_{DS}$ increases. At the high $V_{DS}$, however, $I_{DS}$ may increase rapidly due to the immediately electron injection from $L_R$ to $L_D$ (green arrow).

In summary, we have established a distinct photoconductor based on two separated dye-sensitized solar cells using the common CE. A transistor-like performance is achieved with the photo-gated, showing the promising for potentials such as a logic element in optical communication system. The characteristic transport behaviors observed are completely understandable in view of the electric potential level probably existing in the present device.

**Acknowledgement:**

The authors thank Dr. Zaiping Guo and Dr. Rong Zeng at the ISEM of the University of Wollongong (Australia) for their helpful discussions. This work is partly sponsored by the Science and Technology Commission of Shanghai Municipality (Nano-project No. 0752nm017), the National Natural Science Foundation of China (No. 5062057 and 10774098), the National Science Foundation of China, the Ministry of Science and Technology of China (973 Projects, No. 2006CB601005), and Shanghai Leading Academic Discipline Project (No. S30105).

**References Captions:**

FIGURE CAPTIONS:

Fig. 1.

Schematic diagrams for the preparation and architecture of the present device cell: (a) etching method using a needle. (b) device structure consisting of the counter electrode (CE) opposite to the source (S) and drain (D).

Fig. 2

Current density $J_{DS}$ vs. bias voltage $V_{DS}$ for the present device in the dark (off state) and under illumination (on state). The inset is the calculated differential conductance.

Fig. 3

Characteristic transport loops for $V$-$I$ (left axis) and $V$-$J$ (right axis) for the present device. The inset is a schematic diagram showing the effective dyed-TiO2 area of $0.14 \times 0.5$ cm$^2$ on source and $0.36 \times 0.5$ cm$^2$ on drain, respectively.

Fig. 4

Photo-excited voltage $V_{CE\text{-}D}$ vs. bias voltage $V_{DS}$ for the characterization of the difference between $L_D$ and $L_R$: (a) $V_{CE\text{-}D}$ vs. $V_{DS}$ examined during the $V_{DS}$-$I_{DS}$ loop measurement. The rectangle frame notes the part corresponding to $V_{DS}<0$. (b) Enlarged rectangle frame mentioned above. It is replotted partly from Fig. 5(a) as $V_{CE\text{-}D}$ (left axis) and partly from Fig. 4 as $I_{DS}$ (right axis).

Fig. 5

Schematic diagram for the various potential levels involving in the present device cells. The solid lines represent the dominant direction of electron flowing, and the dash lines represent the potential direction of electron movement.

**Fig.1**

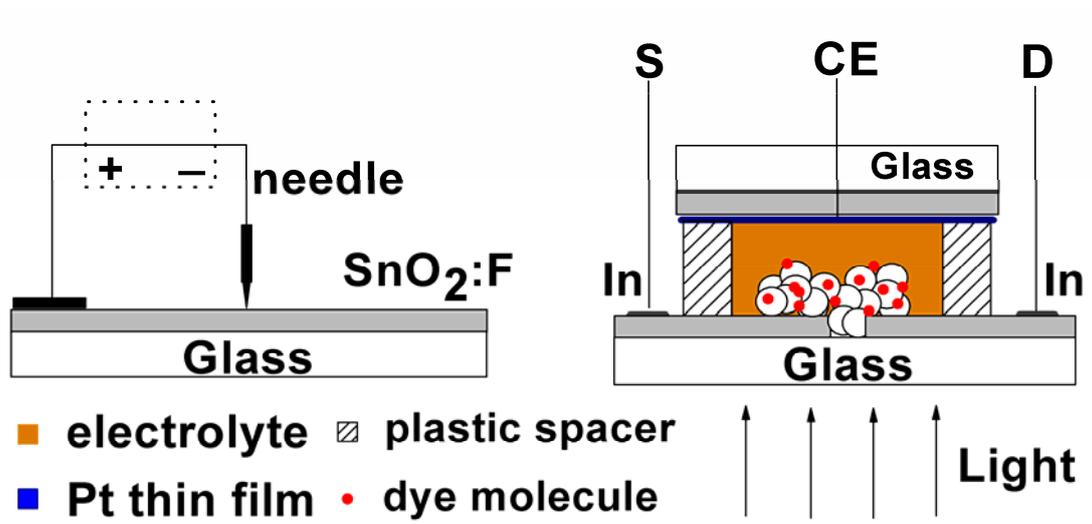

**Fig.2**

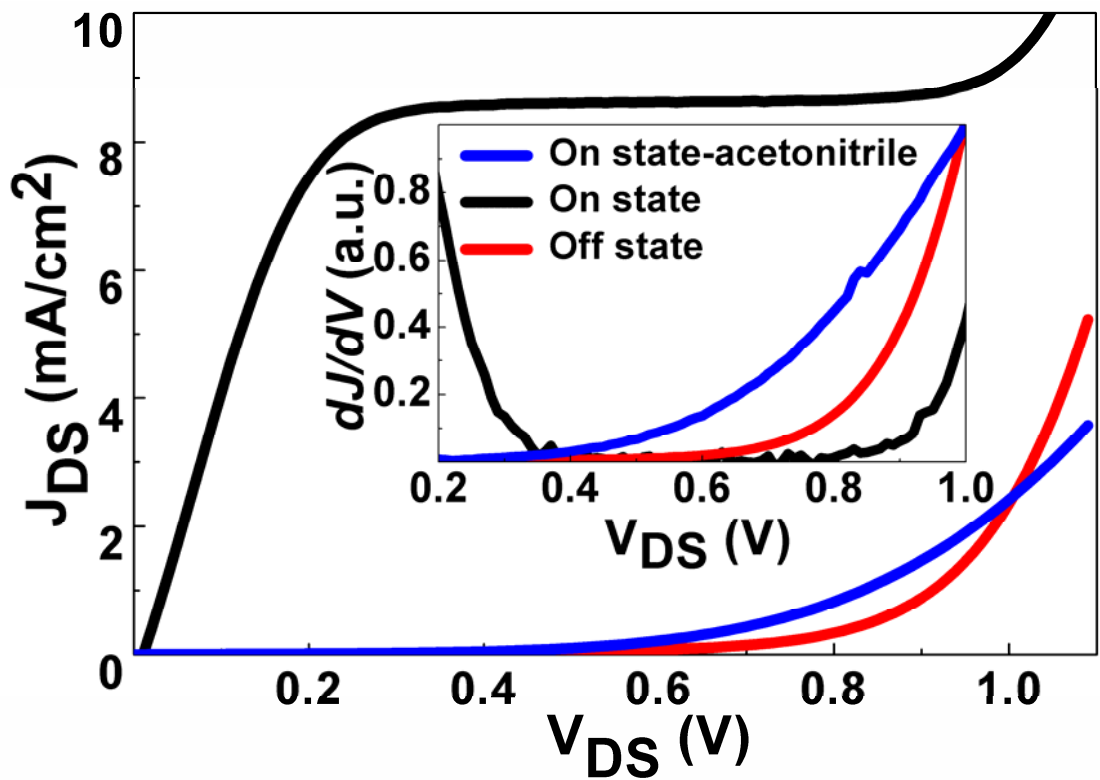

**Fig. 3**

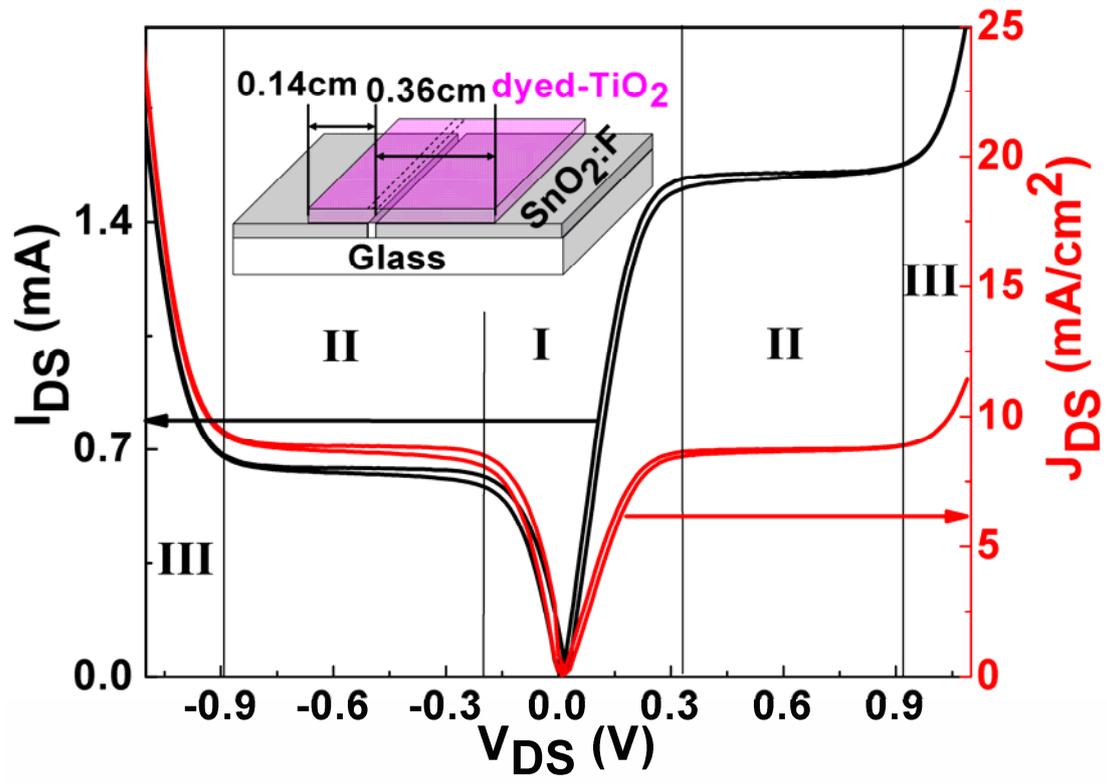

**Fig. 4**

**(a)**

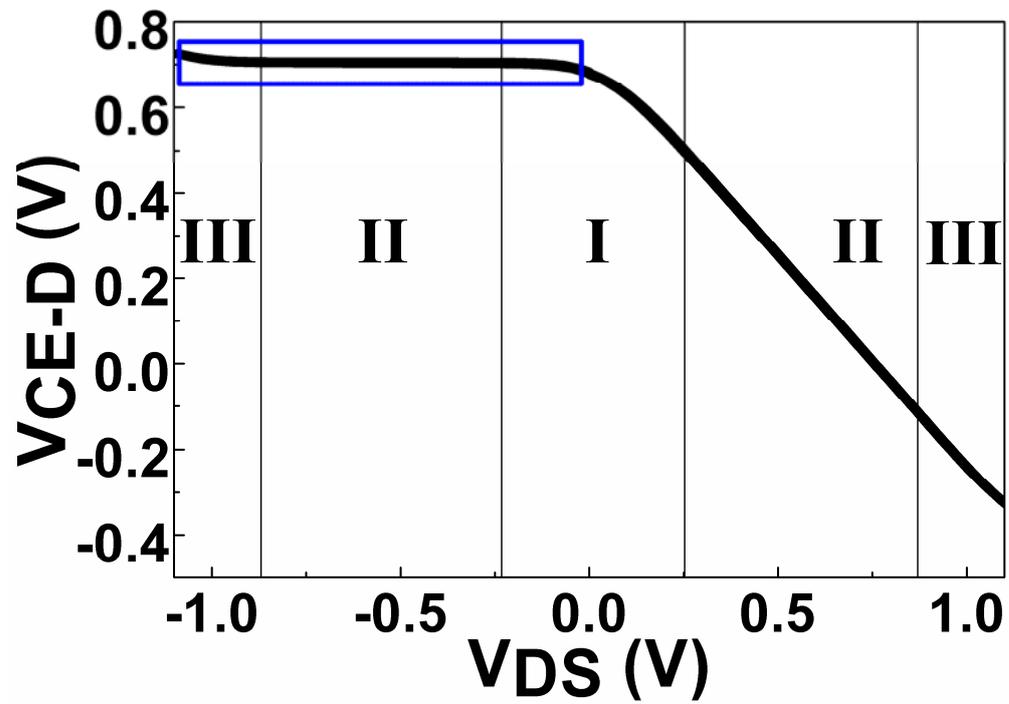

**(b)**

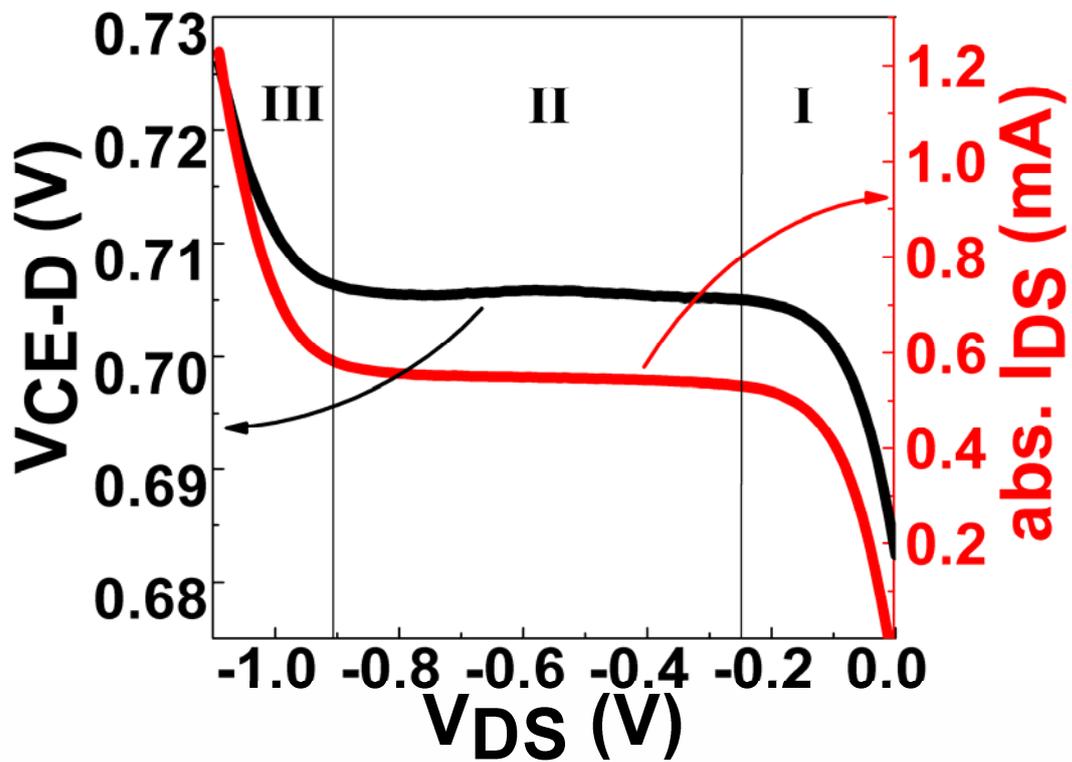

**Fig. 5**

Fig. 5